\documentclass[twocolumn,floatfix,showpacs,nofootinbib]{revtex4}
\usepackage{graphicx}
\usepackage{amsmath}
\usepackage{amssymb}
\usepackage{bm}
\begin{document}

\title{Numerical test of the theory of pseudo-diffusive transmission at the Dirac point of a photonic band structure}
\author{R. A. Sepkhanov}
\affiliation{Instituut-Lorentz, Universiteit Leiden, P.O. Box 9506, 2300 RA Leiden, The Netherlands}
\author{C. W. J. Beenakker}
\affiliation{Instituut-Lorentz, Universiteit Leiden, P.O. Box 9506, 2300 RA Leiden, The Netherlands}
\date{December 2007}
\begin{abstract}
It has recently been predicted that a conical singularity (= Dirac point) in the band structure of a photonic crystal produces an unusual $1/L$ scaling of the photon flux transmitted through a slab of thickness $L$. This inverse-linear scaling is unusual, because it is characteristic of radiative transport via diffusion modes through a disordered medium --- while here it appears for propagation of Bloch modes in an ideal crystal without any disorder. We present a quantitative numerical test of the predicted scaling, by calculating the scattering of transverse-electric (TE) modes by a two-dimensional triangular lattice of dielectric rods in air. We verify the $1/L$ scaling and show that the slope differs by less than 10\% from the value predicted for maximal coupling of the Bloch modes in the photonic crystal to the plane waves in free space. 
\end{abstract}
\maketitle

\begin{figure}[tb]
\centerline{\includegraphics[width=0.9\linewidth]{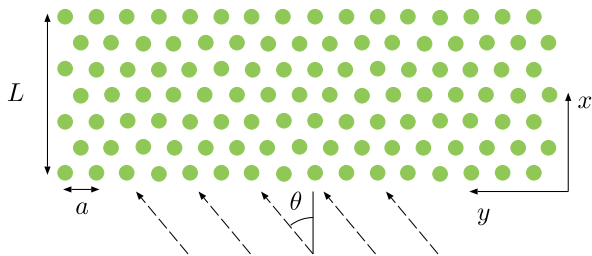}}
\caption{\label{layout_rods}
Top view of a two-dimensional photonic crystal formed by dielectric rods on a triangular lattice in the $x-y$ plane, aligned along the $z$-direction. The lattice constant $a$ (centre-to-centre distance of the rods) is indicated. We calculate the transmission through the slab of thickness $L$ of radiation incident near the $K$-point of the photonic crystal, and find that it scales as $1/L$. 
}
\end{figure}

\begin{figure}[tb]
\centerline{\includegraphics[width=0.9\linewidth]{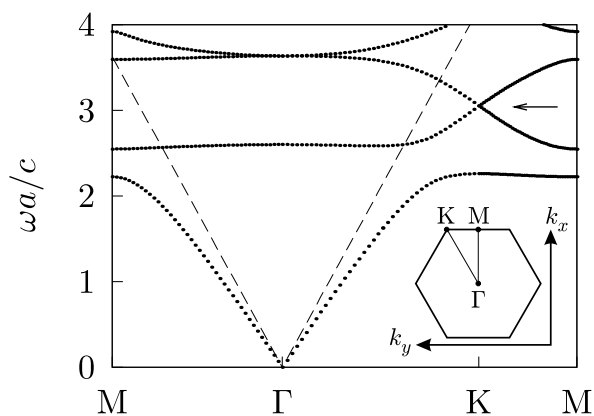}}
\caption{\label{band_rods_2}
Electromagnetic band structure of the photonic crystal shown in Fig.~\ref{layout_rods}, calculated for a dielectric constant $\epsilon=14$ inside the rods and $\epsilon=1$ (air, speed of light $c$) outside the rods. (We used the {\sc mpb} software package for this type of calculation~\cite{Joh01}.) The rods (radius $r=0.27\,a$) occupy a fraction $f=0.26$ of space in the crystal. The bands are shown for the case that the magnetic field is parallel to the rods (TE modes). The arrow points  to the conical singularity (Dirac point) and the dashed line shows the dispersion relation in free space. The first Brillouin zone is drawn in the inset. (Note that the $\Gamma-M$ direction is perpendicular to the $x=0$ interface of the photonic crystal, for the orientation of Fig.~\ref{layout_rods}.)}
\end{figure}

Two-dimensional photonic crystals with a triangular lattice (such as shown in  Fig.~\ref{layout_rods}) have been studied extensively~\cite{Pli91,Vil92,Sak95,Sus95,Joa95,Cas96,Not00,Fot03,Ber04,Guv04,Par04,Mou05,Gaj06}, in particular because they have a well-developed band gap. For frequencies inside this gap the transmission through the crystal decays exponentially with the thickness $L$. The band structure has another interesting feature which has received much less attention, namely the conical singularity that appears at the corner (= $K$-point) of the hexagonal first Brillouin zone~\cite{Pli91}. As indicated in Fig.~\ref{band_rods_2}, at a given wave vector near the $K$-point two Bloch modes are nearly degenerate in frequency. The envelopes of the Bloch modes satisfy a pair of coupled differential equations that have the same form as the Dirac equation of relativistic quantum mechanics~\cite{Rag06}. Hence the name ``Dirac point'' given to the conical singularity. The essential difference between a band gap and a Dirac point is that the density of states is zero for a finite frequency interval in the former case, but only at a single frequency in the latter case.

Motivated by an electronic analogue (graphene~\cite{Two06}), Bazaliy and the authors~\cite{Sep07} have recently predicted a new signature of the conical singularity: near the Dirac point the photon flux $I$ transmitted through a slab of photonic crystal is predicted to scale as $1/L$ with the thickness $L$ of the slab. The $1/L$ scaling is called ``pseudo-diffusive'' due to its reminiscence of diffusion through a disordered medium --- although here it appears for Bloch modes in the absence of any disorder inside the photonic crystal. 

More quantitatively, the prediction of Ref.~\cite{Sep07} is that at the Dirac point
\begin{equation}
I=I_{0}\Gamma_{0}\frac{1}{L},\;\;0<\Gamma_{0}<1/\pi,\label{ItotalDirac}
\end{equation}
with $I_{0}$ the incident photon current per transverse mode and $I$ the transmitted photon flux (= transmitted photon current per unit width). The coefficient $\Gamma_{0}$ that determines the slope of the $1/L$ scaling depends on the coupling strength of the Bloch modes  inside the photonic crystal to the plane waves outside. For maximal coupling one has $\Gamma_{0}=1/\pi$~\cite{Two06,Sep07}.

It is the purpose of this paper to test the prediction of Ref.~\cite{Sep07} quantitatively, by means of a numerical solution of the scattering problem. (An independent test in Ref.~\cite{Zha07} provides only a qualitative comparison.) By means of an exact solution of Maxwell's equations we can test how well the Dirac equation used in Ref.~\cite{Sep07} describes the scattering near the Dirac point. Furthermore, we can determine the slope $\Gamma_{0}$ --- which is beyond the reach of the Dirac equation and was left undetermined in Ref.~\cite{Sep07}. 

We solve the scattering problem in the geometry of Fig.~\ref{layout_rods} for the parameters listed in Fig.~\ref{band_rods_2}. The transmitted photon flux for a given incident plane wave $\propto e^{ik_{x}x+ik_{y}y}$ is calculated as a function of frequency $\omega=c\sqrt{k_{x}^{2}+k_{y}^{2}}$ for a given thickness $L$ of the crystal. (The transverse width is infinite in the calculation.) We use the finite-difference time-domain method~\cite{Taf05}, as implemented in the {\sc meep} software package~\cite{Far06}.

\begin{figure}[tb]
\centerline{\includegraphics[width=0.6\linewidth]{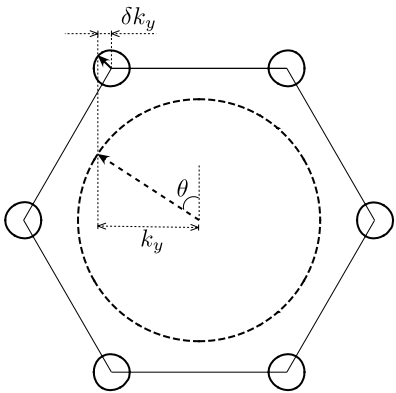}}
\caption{\label{EFC}
Equifrequency contours for the photonic crystal of Fig.\ \ref{band_rods_2}, calculated for $\omega=2.89\,c/a=0.95\,\omega_{D}$. The contours (thick solid lines) are centered at the corners of the first Brillouin zone, and are approximately circular with a slight trigonal distortion. The dashed circle is the equifrequency contour in free space, at the same $\omega$. An incident plane wave at an angle $\theta$ (dashed arrow) is coupled to Bloch modes in the crystal with the same wave vector component $k_{y}$ (solid arrow shows wave vector of the envelope field). When $\omega\rightarrow\omega_{D}$, the radius of the equifrequency contours shrinks to zero and the incident plane wave can only couple to evanescent (exponentially decaying) Bloch modes.}
\end{figure}

To make contact with Ref.~\cite{Sep07} we first extract from Fig.~\ref{band_rods_2} the parameters $\omega_{D}=3.05\,c/a$, $v_{D}=0.369\,c$ that characterise the conical singularity in the band structure,
\begin{equation}
\delta\omega\equiv\omega-\omega_{D}=v_{D}|\delta\bm{k}|.\label{omegakrelation}
\end{equation}
Here $\delta{\bm k}=\bm{k}-\bm{K}$ is the displacement of the wave vector $\bm{k}$ from the $K$-point, with wave vector $\bm{K}=\frac{2}{3}\pi a^{-1}(\sqrt{3},1)$. The velocity $v_{D}$ is the group velocity of Bloch modes at frequencies near the frequency $\omega_{D}$ of the Dirac point. A given $\delta k_{y}$ corresponds to an angle of incidence 
\begin{equation}
\theta=\arcsin{\left[\frac{c}{\omega}(K_{y}+\delta k_{y})\right]}.\label{thetadef}
\end{equation}
In particular, $\delta k_{y}=0$ and $\omega=\omega_{D}$ correspond to $\theta=\arcsin{\left(2\pi c/3\omega_{D}a\right)}\equiv\theta_{0}$. For our parameters $\theta_{0}=43^{\circ}$.

As indicated in Fig.\ \ref{EFC}, an incident plane wave couples to Bloch modes in the photonic crystal with the same $k_{y}$. Propagating envelope modes have wave vector on the equifrequency contour centered at a $K$-point. As the frequency $\omega$ approaches the Dirac frequency $\omega_{D}$, the radius of the equifrequency contour shrinks to zero, and the incident plane wave can only couple to evanescent modes. These decay exponentially away from the interface, with a decay length $\propto 1/|\delta k_{y}|$ which becomes infinitely long at the $K$-point. 

The crucial difference between transmission at the Dirac frequency and inside a band gap is this: In both cases, the photonic crystal supports only evanescent Bloch modes, but inside the band gap the decay length as a function of angle of incidence has a finite maximum value  --- while at the Dirac frequency the maximum decay length is infinite. As a consequence, angular averaging of the transmitted intensity over some narrow range of incident angles around $\theta_{0}$ gives an exponentially decaying transmission inside the band gap, but only an algebraic $1/L$ decay at the Dirac frequency \cite{Sep07}.

For a quantitative description of this scaling behavior we need to consider the coupling strength of the Bloch modes inside the crystal to the plane waves outside. The transfer matrix of the interface at $x=0$ and $x=L$, which determines this coupling, is characterised by two parameters $\beta$ and $\gamma$. These parameters enter into the expression for the transmission probability $T(\delta k_{y},\delta\omega)$, which is defined as the ratio of transmitted to incident photon flux for an incident plane wave [frequency $\omega=\omega_{D}+\delta\omega$ and angle of incidence $\theta$ related to $\delta k_{y}$ by Eq.\ \eqref{thetadef}]. The result is~\cite{Sep07}
\begin{align}
\frac{1}{T}={}&\left(\frac{\delta\omega L\sin \sigma}{v_{D}\sigma}\cosh 2\beta-\cos \sigma\sinh 2\beta\sinh 2\gamma\right.\nonumber\\
&\left.\mbox{}-\frac{\delta k_{y}L\sin \sigma}{\sigma}\sinh 2\beta\cosh 2\gamma\right)^{2}\nonumber\\
&+\left(\cos \sigma\cosh 2\gamma+\frac{\delta k_{y}L\sin \sigma}{\sigma}\sinh 2\gamma\right)^{2},\label{Tresult}
\end{align}
with $\sigma=L\sqrt{(\delta\omega/v_{D})^{2}-\delta k_{y}^{2}}$.

\begin{figure}[tb]
\centerline{\includegraphics[width=0.9\linewidth]{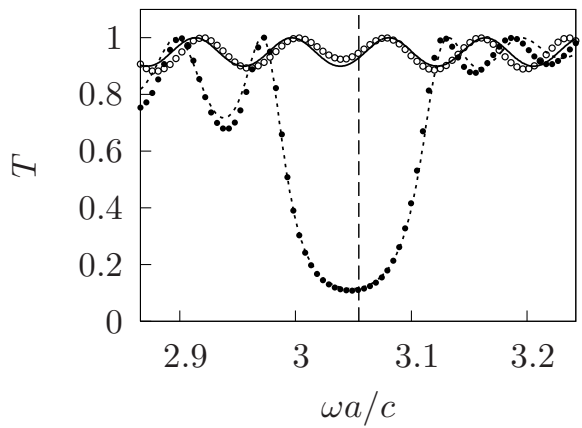}}
\caption{\label{modes1}
Transmission probability through the slab of photonic crystal of thickness $L=8\sqrt{3}\,a$. The data points are the numerical results, the curves are calculated from Eq.~\eqref{Tresult} with the interface parameters of Eq.~\eqref{betagamma}. The vertical dashed line indicates the Dirac frequency $\omega_{D}$. This plot is for a single incident plane wave with $\delta k_{y}=0$ (open data points, solid curve) and $\delta k_{y}=-(\pi/30)a^{-1}$ (filled data points, dotted curve).
}
\end{figure}

We extract the two interface parameters
\begin{equation}
\beta=-0.094,\;\;\gamma=-0.133\label{betagamma}
\end{equation}
from the $T$ versus $\delta\omega$ dependence at $\delta k_{y}=0$, plotted in Fig.~\ref{modes1}. In the same figure we show that the $\delta k_{y}$ dependence of these parameters is weak for $\delta k_{y}a\ll 1$, as was assumed in Ref.~\cite{Sep07}, since the same set of parameters~\eqref{betagamma} also describes the $\delta\omega$ dependence of $T$ at nonzero $\delta k_{y}$.

\begin{figure}[tb]
\centerline{\includegraphics[width=0.9\linewidth]{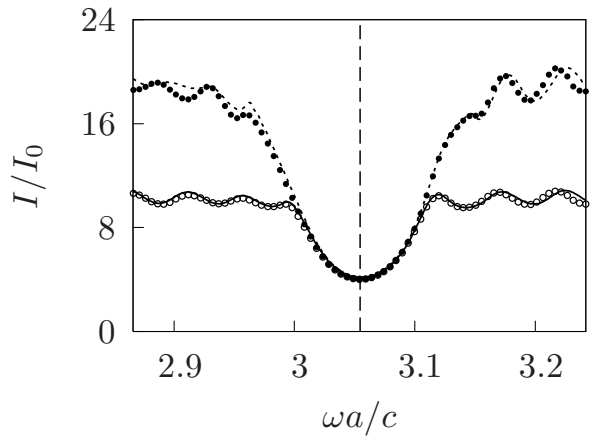}}
\caption{\label{modes2}
Transmitted flux~\eqref{Idef} for $L=13\sqrt{3}\,a$. Data points are the numerical results, curves are calculated from Eq.~\eqref{Tresult}. The vertical dashed line indicates the Dirac frequency $\omega_{D}$. This plot is for a range $|\delta k_{y}|\leq\Delta$ of incident wave vectors, with $\Delta=(\pi/30)a^{-1}$ for the open data points and solid curve; $\Delta=(\pi/15)a^{-1}$ for the filled data points and dotted curve.
}
\end{figure}

To test for the $1/L$ scaling we need to consider a range $-\Delta<\delta k_{y}<\Delta$ of incident transverse wave vectors. (This corresponds to a range $\Delta\theta\simeq2c\Delta/\omega_{D}\cos{\theta_{0}}$ of incident angles centered at $\theta_{0}$.) According to Ref.~\cite{Sep07} the $1/L$ scaling is reached when $L\gtrsim 1/\Delta$. We calculate the transmitted photon flux $I(\omega)$ in this range of wave vectors,
\begin{equation}
I(\omega)=I_{0}\int_{-\Delta}^{\Delta}\frac{\delta k_{y}}{2\pi}\,T(\delta k_{y},\delta\omega=\omega-\omega_{D}).\label{Idef}
\end{equation}
As shown in Fig.~\ref{modes2}, we find a strong dependence of $I$ on the range of wave vectors $\Delta$ away from the Dirac frequency --- but not at the Dirac frequency, where the transmitted flux reaches a minimum $I_{\rm min}$ which is $\Delta$ independent for $\Delta\gtrsim 1/L$.\footnote{
The frequency $\omega_{\rm min}$ of the transmission minimum is slightly offset from the Dirac frequency $\omega_{D}$, but the relative offset is small and vanishes with increasing $L$: $|\omega_{\rm min}-\omega_{D}|/\omega_{D}\approx 10^{-2}\,a/L$. We have checked that it makes no difference for the $1/L$ scaling whether we calculate the transmitted flux at $\omega_{\rm min}$ or at $\omega_{D}$.}   

\begin{figure}[tb]
\centerline{\includegraphics[width=0.9\linewidth]{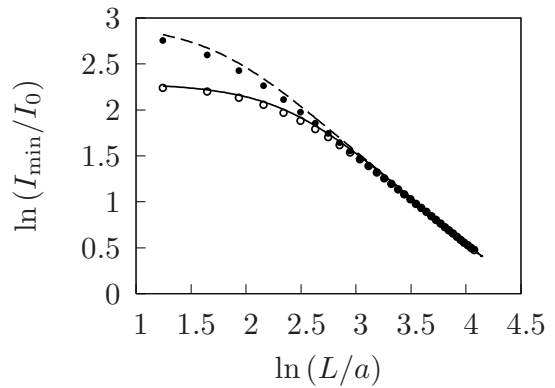}}
\caption{\label{modes3}
Transmitted flux $I_{\rm min}$ at the minimum near the Dirac point versus the thickness $L$ of the slab. Open data points are for $\Delta=(\pi/30)a^{-1}$, filled data points are for $\Delta=(\pi/15)a^{-1}$. The solid and dashed lines show the analytical prediction from Eq.~\eqref{Tresult} with the interface parameters of Eq.~\eqref{betagamma}. 
}
\end{figure}
\begin{figure}[tb]
\centerline{\includegraphics[width=0.9\linewidth]{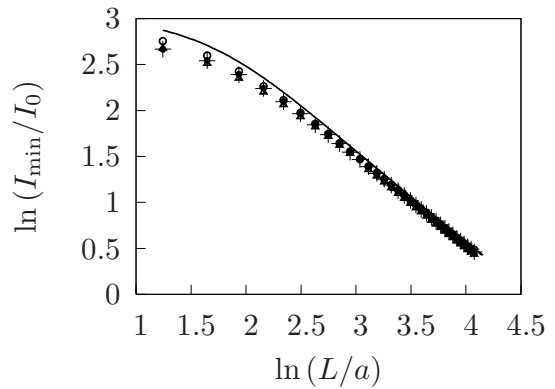}}
\caption{\label{Id}
Minimal transmitted flux versus slab thickness for the four data sets tabulated in Table~\ref{table}. We took $\Delta=(\pi/15)a^{-1}$ in each case. The analytical result for maximal coupling ($\beta=\gamma=0$) is indicated by the solid curve.
}
\end{figure}

In Fig.~\ref{modes3} we plot the $L$ dependence of $I_{\rm min}$ on a double-logarithmic scale. For $L\gg\Delta^{-1}$ the predicted $1/L$ scaling of Eq.~\eqref{ItotalDirac} is obtained, with a coefficient $\Gamma_{0}=0.30$. This coefficient is just 6\% smaller than the value $\Gamma_{0}=1/\pi$ reached for maximal coupling of Bloch modes and plane waves at the interfaces between the photonic crystal and free space.

\begin{table}[h]
\begin{tabular}{|c|c|c|c|c|c|c|c|}

\hline
                   & $\epsilon$ & $f$   & $\omega_{D}c/a$ & $v_{D}/c$ & $\beta$ & $\gamma$ & $\Gamma_{0}$ \\
\hline
$\bm{\circ}$       & 14         & 0.26  & 3.05            & 0.369      & -0.094  & -0.133  & 0.298        \\
\hline
$\blacktriangle$   & 14         & 0.43  & 2.50            & 0.254      & 0.065   & -0.162  & 0.295        \\
\hline
$\bullet$          & 8.9        & 0.33  & 3.03            & 0.432      & -0.095  & -0.197  & 0.298        \\
\hline
$\bm{+}$           & 8.9        & 0.40  & 2.83            & 0.393      & -0.045  & -0.199  & 0.298        \\
\hline
\end{tabular}
\caption{\label{table}
Parameters representing four different triangular lattice photonic crystals. Symbols on the left correspond to the data points in Fig.~\ref{Id}.
}
\end{table}

To investigate how generic these results are, we have repeated the calculation for different values of the dielectric constant $\epsilon$ of the rods and for different filling fractions $f$ (related to the radius $r$ of the rods by $f=2\pi r^2/\sqrt{3} a^2$). The parameters corresponding to the four sets of data are tabulated in Table~\ref{table}. In Fig.~\ref{Id} we show the $L$ dependence of the minimal transmitted flux for each data set. In each case we find $1/L$ scaling with a slope $\Gamma_{0}$ that remains within 8\% of the maximal value $\Gamma_{0}=1/\pi$.

In conclusion, we have presented a quantitative numerical test of the
applicability of the Dirac equation~\cite{Rag06} to a photonic crystal
with a conical singularity in the band structure. The numerical results
are in good agreement with the analytical predictions~\cite{Sep07} for
the transmission through a finite slab. In particular, our numerical
calculation demonstrates the $1/L$ scaling of the transmitted photon flux with a slope that
is close to the value for maximal coupling at the interface with free
space. This finding implies that transmission experiments can be used to
search for intrinsic properties of the Dirac point in the band
structure, not hindered by a weak coupling to the outside.

\acknowledgments
We have benefited from discussions with M. de Dood. This research was supported by the Dutch Science Foundation NWO/FOM.

\end{document}